# Coupled high Q-factor Surface Nanoscale Axial Photonics (SNAP) microresonators

M. Sumetsky, K. Abedin, D. J. DiGiovanni, Y. Dulashko, J.M. Fini, and E. Monberg

*OFS Laboratories, 19 Schoolhouse Road, Somerset, NJ 08873*
*Corresponding author: sumetski@ofsoptics.com*

We experimentally demonstrate series of identical two, three, and five coupled high Q-factor Surface Nanoscale Axial Photonics (SNAP) microresonators formed by periodic nanoscale variation of the optical fiber radius. These microresonators are fabricated with a 100 μm period along an 18 μm radius optical fiber. The axial FWHM of these microresonators is 80 μm and their Q-factor exceeds $10^7$. In addition, we demonstrate a SNAP microresonator with the axial FWHM as small as 30 μm and the axial FWHM of the fundamental mode as small as 10 μm. These results may potentially enable the dense integration of record low loss coupled photonic microdevices on the optical fiber platform.

Surface Nanoscale Axial Photonics (SNAP) is a recently introduced platform for fabrication of complex miniature photonics circuits and devices by nanoscale variation of the optical fiber radius [1-3]. It was shown that such dramatically small deformation of the optical fiber (and/or equivalent variation of refractive index) is sufficient to localize the whispering gallery modes (WGMs) propagating along the fiber surface normal to its axis and to create high Q-factor microresonators. Reproducible fabrication of these microresonators demonstrated in [3] supported the robustness of the SNAP platform. Due to the significantly smaller surface roughness of drawn silica compared to the roughness of surfaces fabricated lithographically, it is expected that the SNAP circuits will enable orders of magnitude smaller attenuation of light compared to the planar ring microresonator and photonic crystal microcavity circuits [4-6].

Individual high Q-factor SNAP microresonators have been reported in [1-3]. However, similarly to the photonics circuit platforms developed to date [4-6], the prospective SNAP circuits should include series of microcavities coupled to each other. The idea of coupled ring resonators created at the optical fiber surface has been proposed theoretically [7, 8] but, to date, has not been realized due to significant experimental challenges. This Letter presents the first experimental demonstration of identical coupled SNAP microresonators illustrated in Fig. 1(a).

A SNAP microresonator is a particular case of a bottle microresonator [9] which is formed by nanoscale variation of the optical fiber radius. The typical axial length of a SNAP microresonator can vary from several microns to several millimeters. In our experiments, microresonators were created along a specially fabricated 18 μm radius silica optical fiber by local annealing with a focused $CO_2$ laser beam [3]. The introduced nanoscale variation of the fiber radius and its refractive index is explained by relaxation of the tension with which the fiber was drawn [10, 11]. In agreement with [10, 11], the fiber radius increases with the heating power achieving saturation at the radius variation of the order of ten nanometers. The saturation corresponds to the condition of tension relaxation. In the first experiment of this paper, the coupled SNAP microresonators are fabricated by localized heating below the saturation. An increase of the beam power above the saturation value causes a decrease in the fiber radius, which can be also controlled at a nanoscale. This effect is used in our second experiment, which demonstrates fabrication of two identical coupled microresonators with a single beam exposure. Finally, in the third experiment of this Letter, exploiting the beam power over the saturation value, we demonstrate a method of fabrication of SNAP microresonators with dimensions of a few tens of microns.

In our first experiment, we fabricated a single microresonator and the series of two, three and five coupled microresonators (Fig. 1(b)-(e)). The microresonator spectrum and radius variation were determined using the method of a scanning optical microfiber (MF) illustrated in Fig. 1(a) [12, 13, 2]. An MF (a micron diameter waist of a biconical fiber taper connected to the JDSU tunable laser source and detector, 3 pm spectral resolution) was aligned normal to the test fiber. The MF was translated along the test fiber and touched it periodically at the contact points, where the WGM resonant transmission spectrum was measured. The series of transmission spectra shrunk along the vertical axes, which are measured at the points spaced by 10 μm, are shown in Fig. 1(b)-(e). The effective radius variation, $\Delta r_{eff}(z) = \Delta r(z) + \Delta n(z) r_0 / n_f$, which summarizes the change $\Delta r(z)$ in physical radius $r_0$ and the change $\Delta n(z)$ in refractive index $n_f$ of the fiber [2, 3], was determined by enveloping the resonance regions (bold curves in Fig. 1 (b)-(e)). The single SNAP microresonator in Fig. 3(b) has the radius variation height of 4 nm and the axial FWHM of 80 μm. The microresonator effective radius variation, $\Delta r_{eff}(z)$, forms a quantum well [1, 2] which has four axial states indicated by resonant peaks in the transmission spectra. Fig. 1(c) demonstrates two coupled bottle microresonators created by two similar $CO_2$ laser exposures spaced by 100 μm (bold curve in Fig. 1(c)). In terms of quantum mechanics [1, 2], these resonators form a double quantum well structure. The coupling of the lowest (fundamental) axial states and splitting of the corresponding resonances is not resolved in Fig. 1(c) because the potential barrier separating these states is too large. However, the subsequent resonances correspond to a smaller barrier separation and their splitting into doublets is well resolved. Similarly, for the three and five microresonators shown in Fig. 1(d) and (e), these resonances are split into triplets and pentets. These adjacent multiplet resonances are separated from other resonances by band gaps. In agreement with the theory [2], the Q-factor of the fabricated microresonators decreases with an increase in the intensity of light at the point of contact with the MF. Thus, the Q-factor maxima are achieved in the evanescent regions and at the nodes of the localized states. In these regions, the Q-factor was measured with a Tunics external cavity tunable laser (0.2 pm resolution) and found to be greater than $10^7$. A more accurate measurement of the Q-factor of much less shallow bottle microresonators showed that it can exceed $10^8$ [14], whereas the Q-factor of silica WGM resonators can be as large as $10^{10}$ [15].

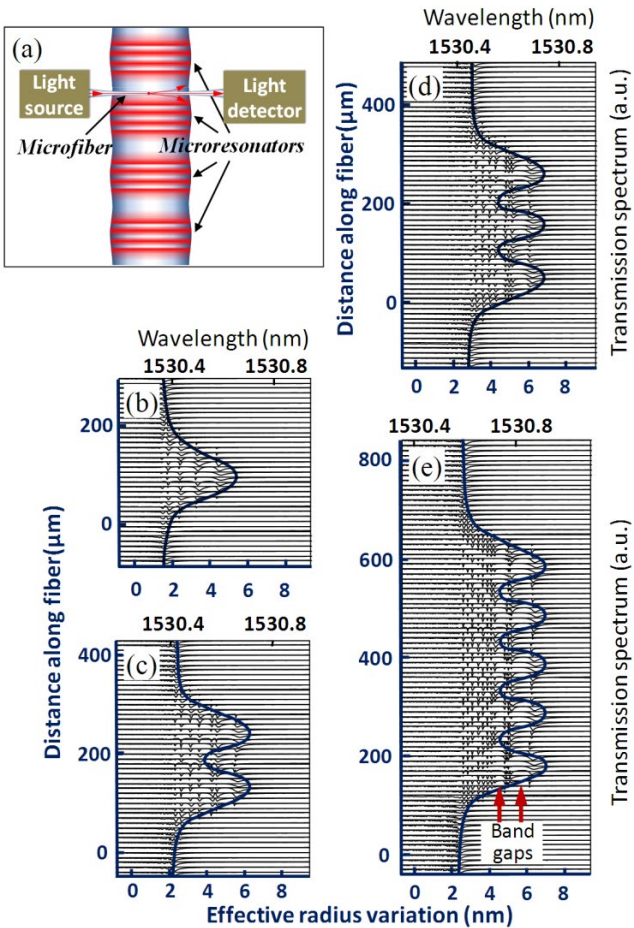

Fig. 1. (a) – Illustration of coupled SNAP microresonators excited by an .MF attached to a SNAP fiber. (b) – Characterization of an individual microresonator. (c), (d) and (e) – Characterization of two, three, and five coupled microresonators, respectively. In figures (b), (c), (d) and (e), multiple plots spaced by 10 μm and shrunk along the vertical axis are the resonance spectra measured along the test fiber at points spaced by 10 μm. Bold curves represent the radius variation of microresonators.

Theoretically, the WGM behavior in a SNAP fiber is described by the Schrödinger equation with the potential proportional to the effective radius variation $\Delta r_{eff}(z)$ and the energy proportional to the wavelength variation [1, 2].

To estimate the resonance splitting, the radius variation near the bottom of quantum wells was approximated by the quadratic dependence $\Delta r_{eff}(z)=-(z-z_w)^2/(2R_w)$ with the axial curvature radius $R_w$ of the bottle microresonator. The fiber shape in the barrier region was approximated by a similar dependence $\Delta r_{eff}(z)=(z-z_b)^2/(2R_b)$. Then, using the known formula for the splitting of levels in a double quantum well [16] we find the resonance wavelength splitting for two coupled microresonators:

$$\delta=(2\pi n_f)^{-1}\lambda_r^2(r_0 R_w)^{-1/2}\exp[-2\pi^2 n_f \Delta\lambda\lambda_r^{-2}(r_0 R_b)^{1/2}]. \quad (1)$$

From the experimental data in Fig. 3(c), the parameters of Eq. (1) are determined as follows: the refractive index of silica $n_f=1.46$, fiber radius $r_0=18$ μm, the resonance wavelength $\lambda_r=1.53$ μm, the deviation of $\lambda_r$ from the top of the barrier $\Delta\lambda=0.05$ nm, and the bottle and barrier axial radii $R_w=R_b=0.2$ m. With these parameters, Eq. (1) yields $\delta=0.04$ nm which is in a good agreement with the data depicted in Fig. 3(c). In accordance with the theory [8], the characteristic width $\delta$ of the multiplet resonance bands in Fig. 3(c), (d) and (e) is independent of the number of coupling microcavities.

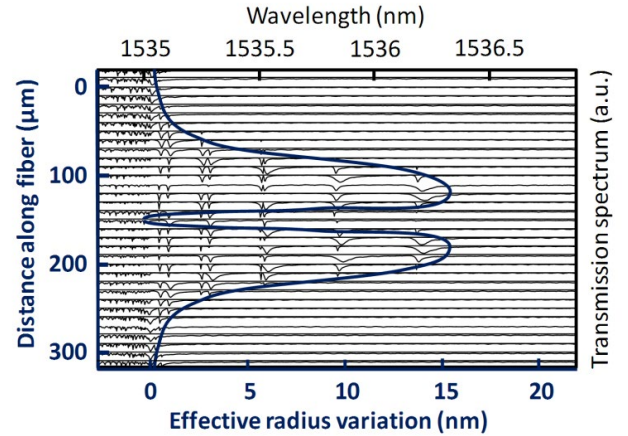

Fig. 2. Characterization of a double microresonator (double quantum well) structure fabricated by a single oversaturated beam exposure (the definition of plots is similar to Fig. 1).

Another method to create coupled SNAP microresonators is demonstrated in our second experiment. We have found that, in agreement with the results on the optical fiber elongation with temperature [10], increasing the heating power of the $CO_2$ laser beam beyond the saturation value [3] leads to nanoscale shrinkage of the fiber radius. As an example, Fig. 2 shows a double microresonator (double quantum well) structure created with a single beam exposure. The reason for the radius variation profile shown in Fig. 2 is explained as follows. The central part of the beam, characterized by a higher power, heats the fiber to a higher temperature (larger than the saturation value), which causes reduction of the fiber radius in the central region. Notice that the barrier separating quantum wells in Fig. 2 has a surprisingly small FWHM equal to 20 μm only.

Thus, using the laser beam which, in its central region, heats the fiber to the temperature above the saturation value, allowed us to introduce narrow potential barriers and significantly reduce the characteristic axial length of the effective radius variation. In our third experiment, which is based on the above idea, a SNAP microresonator with a record small axial dimension was created between two barriers. The barriers were formed by two exposures with oversaturated similar beams spaced by 50 μm. For a better axial resolution, the introduced radius variation profile shown in Fig. 3 is characterized by resonant spectra measured at the points spaced by 5 μm (rather than 10 μm as in the previous experiments). Two side microresonators in Fig. 3 correspond to the regions of undersaturated heating, similar to those in Fig. 2. Increasing the power of the laser beams allowed us to move the tops of the barriers into the continuous spectrum of the SNAP fiber (Fig. 3). The created microresonator has the record small FWHM of 30 μm. The free spectral range (FSR) of this microresonator is 1 nm, which is more than 10 times greater than the FSR of the microresonator shown in Fig. 1(b). The axial FWHM of the fundamental mode of this microresonator is 10 μm. Similar to the microresonators depicted in Figs. 1 and 2, the width of resonances in Fig. 3 depends on the MF position and is proportional to the local field intensity. For example, the field distribution of the fundamental mode in Fig. 3 has a single maximum; the next state has two maxima and a node in the middle, etc. (wavy bold curves in Fig. 3).

Similarly, applying more than two equally spaced oversaturated exposures, it is possible to create the series of coupled microresonators. As opposed to the coupled microresonators shown in Fig. 1, these microresonators will be situated in the continuous rather than discrete spectrum of the SNAP fiber. For this reason, these microresonators can be directly coupled to free WGMs propagating along the fiber. Experimental and theoretical consideration of this situation is beyond the scope of this Letter.

In summary, it is shown here that the nanoscale variation of the optical fiber radius allows fabrication of the series of identical microresonators coupled to each other. In contrast to other types of coupled photonic microresonators [4-6], the spacing between SNAP microresonators does not include conventional material interfaces and, hence, does not introduce additional attenuation of light. We have demonstrated the series of two, three, and five coupled SNAP microresonators having the 80 μm axial FWHM length and the Q-factor exceeding $10^7$. Using the oversaturated beam exposures, we have demonstrated a SNAP microresonator with the record small axial FWHM of 30 μm and fundamental mode axial FWHM of 10 μm. We believe that these results will potentially enable dense integration of record low loss microdevices including sequences of coupled microresonators on the optical fiber platform.

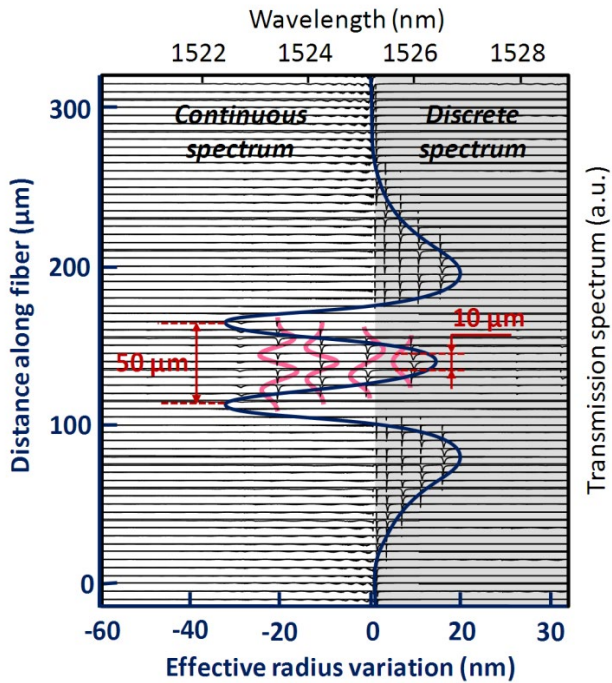

Fig. 3. Characterization of the fiber radius variation introduced by two oversaturated beam exposures (the definition of plots is similar to Fig. 1). The central microresonator (quantum well) belongs to the continuous spectrum of the SNAP fiber. The side quantum wells remain in the discrete spectrum (darken). Bold wavy curves illustrate the field amplitude variation of the modes in the central microresonator.